\documentclass[fleqn,usenatbib]{mnras}
\usepackage{amsmath}
\usepackage{xspace}
\usepackage{mathptmx}
\usepackage{txfonts}
\usepackage{xcolor}

\usepackage[T1]{fontenc}
\usepackage{ae,aecompl}

\usepackage{graphicx}	% Including figure files
\usepackage{amssymb}	% Extra maths symbols
\usepackage{booktabs}
\usepackage{wasysym}
\let\oldhref\href
\renewcommand{\href}[2]{\oldhref{#1}{\hbox{#2}}}

\definecolor{colorl1}{RGB}{0, 51, 153}
\definecolor{colorl2}{RGB}{153, 0, 0}
\definecolor{colorl3}{RGB}{179, 179, 0}
\definecolor{colorl4}{RGB}{51, 102, 0}

\definecolor{colorw1}{RGB}{51, 102, 255}
\definecolor{colorw2}{RGB}{255, 51, 0}
\definecolor{colorw3}{RGB}{255, 214, 51}
\definecolor{colorw4}{RGB}{51, 204, 51}

\newcommand{\hMpc}{{\ifmmode{h^{-1}{\rm Mpc}}\else{$h^{-1}$Mpc}\fi}}
\newcommand{\Mpc}{{\ifmmode{{\rm Mpc}}\else{Mpc}\fi}}
\newcommand{\hkpc}{{\ifmmode{h^{-1}{\rm kpc}}\else{$h^{-1}$kpc}\fi}}
\newcommand{\kpc}{{\ifmmode{ {\rm kpc} }\else{{\rm kpc}}\fi}}
\newcommand{\kms}{{\ifmmode{ {\rm km\,s^{-1}} }\else{ ${\rm km\,s^{-1}}$ }\fi}}
\newcommand{\hMsun}{{\ifmmode{h^{-1}{\rm {M_{\astrosun}}}}\else{$h^{-1}{\rm{M_{\astrosun}}}$}\fi}}
\newcommand{\Msun}{{\ifmmode{{\rm M}_{\astrosun}}\else{${\rm M}_{\astrosun}$}\fi}}
                        
\newcommand{\Mhalo}{{\ifmmode{M_{\rm halo}}\else{$M_{\rm halo}$}\fi}}
\newcommand{\Rvir}{{\ifmmode{R_{\rm vir}}\else{$R_{\rm vir}$}\fi}}
\newcommand{\Mvir}{{\ifmmode{M_{\rm vir}}\else{$M_{\rm vir}$}\fi}}
\newcommand{\Mstar}{{\ifmmode{M_{\rm star}}\else{$M_{\rm star}$}\fi}}
\newcommand{\Vrot}{{\ifmmode{V_{\rm rot}}\else{$V_{\rm rot}$}\fi}}
\newcommand{\ltsima}{$\; \buildrel < \over \sim \;$}
\newcommand{\gtsima}{$\; \buildrel > \over \sim \;$}
\newcommand{\lsim}{\lower.5ex\hbox{\ltsima}}
\newcommand{\gsim}{\lower.5ex\hbox{\gtsima}}

\def\lesssim{\mathrel{\hbox{\rlap{\hbox{\lower4pt\hbox{$\sim$}}}\hbox{$<$}}}}
\def\gtrsim{\mathrel{\hbox{\rlap{\hbox{\lower4pt\hbox{$\sim$}}}\hbox{$>$}}}}

\newcommand{\beq}{\begin{equation}}
\newcommand{\eeq}{\end{equation}}
\def\beqa{\begin{eqnarray}}
\def\eeqa{\end{eqnarray}}
\def\LCDM{\ensuremath{\Lambda}CDM}

\def\head{ \vbox to 0pt{\vss \hbox to 0pt{\hskip 440pt\rm
      LA-UR-10-07069\hss} \vskip 25pt}}

\def \kms {\ifmmode  \,\rm km\,s^{-1} \else $\,\rm km\,s^{-1}  $ \fi }
\def \kpc {\ifmmode  {\,\rm kpc}  \else ${\rm  kpc}$ \fi  }  
\def \hkpc {\ifmmode  {h^{-1}\rm kpc}  \else ${h^{-1}\rm kpc}$ \fi  }  
\def \hMpc {\ifmmode  {h^{-1}\rm Mpc}  \else ${h^{-1}\rm Mpc}$ \fi  }  
\def \Mpch {\ifmmode  {h^{-1}\rm Mpc}  \else ${h^{-1}\rm Mpc}$ \fi  }  
\def \Msun {\ifmmode {\rm M}_{\astrosun} \else ${\rm M}_{\astrosun}$ \fi} 
\def \hMsun {\ifmmode h^{-1}\,\rm M_{\astrosun} \else $h^{-1}\,\rm M_{\astrosun}$ \fi}
\def \Gyr {\ifmmode\, \rm Gyr \else $\,$Gyr \fi}

\def \LCDM {\ifmmode \Lambda{\rm CDM} \else $\Lambda{\rm CDM}$ \fi}
\def \sig8 {\ifmmode \sigma_8 \else $\sigma_8$ \fi} 
\def \OmegaM {\ifmmode \Omega_{\rm m} \else $\Omega_{\rm m}$ \fi} 
\def \Omegab {\ifmmode \Omega_{\rm b} \else $\Omega_{\rm b}$ \fi} 
\def \OmegaL {\ifmmode \Omega_{\rm \Lambda} \else $\Omega_{\rm \Lambda}$\fi} 
\def \Deltavir {\ifmmode \Delta_{\rm vir} \else $\Delta_{\rm vir}$ \fi}
\def \rhocrit {\ifmmode \rho_{\rm crit} \else $\rho_{\rm crit}$ \fi}
\def \rhou {\ifmmode \rho_{\rm u} \else $\rho_{\rm u}$ \fi}
\def \zc {\ifmmode z_{\rm c} \else $z_{\rm c}$ \fi}

\def\Mstar {\ensuremath {M_{*}(<r_{23.5})}~}
\def\r23_5 {\ensuremath {r_{23.5}}~}

\defcitealias{Tollet2016}{NIHAO-IV}

\title[NIHAO XXIII] {NIHAO-XXIII: Dark Matter density shaped by Black Hole feedback}

\author[A.V. Macci\`o et al.]{Andrea V. Macci\`o$^{1,2,3}$\thanks{E-mail: maccio@nyu.edu},
Samuele Crespi$^{1,2}$\thanks{E-mail: sc6459@nyu.edu}, Marvin Blank$^{1,2,4}$, Xi Kang$^5$
\\
$^{1}$New York University Abu Dhabi, PO Box 129188, Abu Dhabi, United Arab Emirates \\
$^2$Center for Astro, Particle and Planetary Physics (CAP$^3$), New York University Abu Dhabi\\
$^3$Max-Planck-Institut f\"ur Astronomie, K\"onigstuhl 17, 69117 Heidelberg, Germany \\
$^4$Institut f\"{u}r Theoretische Physik und Astrophysik, Christian-Albrechts-Universit\"{a}t zu Kiel, Leibnizstr. 15, D-24118 Kiel, Germany\\
$^5$Zhejiang University-Purple Mountain Observatory Joint Research Center for Astronomy, Zhejiang University, Hangzhou, 310027, China
}

\date{Accepted XXX. Received YYY; in original form ZZZ}

\pubyear{2020}

\begin{document}

\label{firstpage}
\pagerange{\pageref{firstpage}--\pageref{lastpage}}
\maketitle
\begin{abstract}

We present a systematic analysis of the reaction of dark matter distribution to galaxy formation
across more than eight orders of magnitude in stellar mass. We extend the previous work presented 
in the NIHAO-IV paper (Tollet et al.) by adding 46 new high resolution simulations of massive galaxies
performed with the inclusion of Black Hole feedback. We show that outflows generated by the AGN
are able to partially counteract the dark matter contraction  due to the large central stellar component in massive haloes. The net effect is to relax the central dark matter distribution that moves to a less cuspy density profiles at halo mass larger than $\approx 3 \times 10^{12}$ \Msun. 
The scatter around the mean value of the density profile slope ($\alpha$) is fairly constant ($\Delta\alpha \approx 0.3$), with the exception of galaxies with halo masses around $10^{12}$ \Msun, at  the transition  from stellar to AGN feedback dominated systems, where the scatter increases by almost a factor three.
We provide useful fitting formulas for the slope of the dark matter density profiles at few percent of the virial radius for the whole stellar mass range: $10^5-10^{12}$ \Msun ($2 \times 10^9-5 \times 10^{13}$ \Msun in halo mass).

\end{abstract}

\begin{keywords}
cosmology: theory -- dark matter -- galaxies: formation -- galaxies: kinematics and dynamics -- methods: numerical
\end{keywords}

%%%%%%%%%%%%%%%%%%%%%%%%%%%%%%%%%%%%%%%%%%%%%%%%%%%
\section{Introduction}\label{sec:introduction}
%%%%%%%%%%%%%%%%%%%%%%%%%%%%%%%%%%%%%%%%%%%%%%%%%%%

A clear picture of the distribution of dark matter in galaxies is one of the 
most important capstones towards a full understanding of the nature of dark matter and 
a firm confirmation of the current structure formation paradigm based on the 
Lambda Cold Dark Matter model \citep[e.g.][]{Planck2014}.

Pure gravity (collision-less) simulations make clear predictions  on the dependence of the  dark matter distribution  on the nature of dark matter constituents \citep[e.g. cold vs. warm][]{navarro_etal97,Maccio2008, Schneider2014, Lovell2012}, and hence provide a tool to shed light on the dark sector; but unfortunately the situation is not so simple. 
In the recent years, it has become more and more evident that the galaxy formation process (dissipational and collisional) is able to alter the dark matter distribution. There are mainly two competing processes at work: the central stellar component acts as a gravitation sink for the dark matter (DM) that then tends to contract towards the center \citep[e.g.][]{Blumenthal86,Gnedin2004,Abadi2010,Schaller2015}, on the other hand vigorous gas outflows generate rapid change of the total gravitational potential and these  fluctuations in the central potential irreversibly \citep[i.e. non-adiabatically][]{Pontzen2012} transfer energy into collision-less particles, thus expanding the DM distribution. 
These strong gas outflows are triggered by stellar (and active galactic nuclei) feedback 
\citep{Navarro1996,Read2005,mashchenko08,Maccio2012,Pontzen2014,Madau2014,Freundlich2020}.

The exact magnitude of the final effect of these two competing effects, 
and their dependence on the adopted schemes and parameters for the description of subgrid processes is still under debate \citep{Dutton2017,Bose2019}, nevertheless there is mounting consensus that baryons do have an important role in shaping the dark matter distribution (see recent review from \citet {Bullock2017} 
and also \citet{Qin2017} for additional effects able to modify the dm distribution).

Several groups have tried to use large sets of high resolution simulations to quantify the impact
of baryons on redistributing dark matter and how this changes as a function of galactic stellar and halo mass.
\cite{DiCintio2014a} using the MaGICC simulation suite \citep{Stinson2013} first showed that 
the modification of the initial DM profile (either leading towards expansion or contraction) is 
linked to the integrated star formation efficiency of a galaxy, which can be
captured by the present day stellar mass–halo mass ratio ($M_{\rm star}/M_{\rm halo}$), and that this result
holds for  different stellar feedback implementations.

\citet{Tollet2016} (hereafter \citetalias{Tollet2016}) has later confirmed and extended the findings of
\cite{DiCintio2014a} using  67 NIHAO galaxies from \cite{Wang2015} covering  halo  mass range of about two decades from $10^{10}$ to $10^{12}$ \Msun. Finally the key role of the ratio  $M_{\rm star}/M_{\rm halo}$ has been  independently confirmed by \cite{Chan2015} using the FIRE simulations from \cite{Hopkins2014} 
\citep[see also][]{Onobe2015}. On the other hand none of the studies mentioned above has been able to go beyond the tip of the star formation efficiency located around the mass of our own Milky Way ($\Mhalo \sim 10^{12}$ \Msun). 
After this mass threshold feedback from supermassive black holes is expected to become the dominant
source of energy in the intra and circum galactic medium \citep[e.g.][]{croton_etal06}.

The effect of AGN feedback on dark matter distribution has been addressed using large scales simulations by, among others, \citet{Duffy2010} using the OWLs simulations and \citet{Peirani2017} using the {\sc horizon} simulations. Both study found an overall contraction of the dark matter, while \cite{Teyssier11} using a single high resolution cluster-size halo, found evidence for slight adiabatic expansion, due to their vigorous AGN feedback.

%and it is then important to see if the trends
%and fitting formulas presented in previous works concerning the reaction of DM to galaxy formation can be extrapolated beyond this mass.

The scope of this Letter is to reanalyze in a consistent fashion the new, enlarged NIHAO sample
with feedback from AGNs,  and to test and extend the previous results from  \citetalias{Tollet2016} to a
much larger stellar mass range in order to assess the effect of BH feedback on the dark matter distribution.
We have a total of 96 galaxies run without AGN feedback (29 more than in the original \citetalias{Tollet2016} paper) and 46 new galaxies run with AGN feedback \citep[from][]{Blank2019}, which allow us to study
the DM reaction to galaxy formation on more than eight orders of magnitude in stellar mass.
We present new fitting formulas able to successfully capture this response as a function of different galaxy parameters (e.g. halo mass, stellar mass etc.) on the whole mass range.

%%%%%%%%%%%%%%%%%%%%%%%%%%%%%%%%%%%%%%%%%%%%%%%%%%%
\section{Simulations} \label{sec:simulations}
%%%%%%%%%%%%%%%%%%%%%%%%%%%%%%%%%%%%%%%%%%%%%%%%%%%

This Letter is based on an extended version of the  NIHAO (Numerical Investigation of Hundred Astrophysical Objects) suite of cosmological hydrodynamical simulations \citep{Wang2015,Blank2019}

The set of simulations are based on the {\sc gasoline2} code \citep{Wadsley2017},
and include compton cooling and photoionisation and heating from the ultraviolet background following \citet{Haardt2012}, metal cooling, chemical enrichment, star formation and feedback from supernovae 
 and massive stars \citep[the so-called Early Stellar Feedback][]{Stinson13}.
The cosmological parameters are set according to \cite{Planck2014}: Hubble parameter
$H_0$= 67.1 \kms Mpc$^{-1}$, matter density $\Omega_\mathrm{m}=0.3175$, dark energy density
$\Omega_{\Lambda}=1-\Omega_\mathrm{m} -\Omega_\mathrm{r}=0.6824$, baryon density
$\Omega_\mathrm{b}=0.0490$, normalization of the power spectrum $\sigma_8 = 0.8344$, 
slope of the inital power spectrum $n=0.9624$, and each galaxy is resolved with at least
half a million elements (dark matter, gas and stars). The mass and spatial resolution vary across the whole sample, from a dark matter particle mass of 
$m_{\rm dm}  = 3.4 \times 10^3$ \Msun (and force softening of $\epsilon =100$  pc) for dwarf galaxies to $m_{\sc dm} = 1.4 \times 10^7$ \Msun and $\epsilon = 1.8$ kpc for the most massive galaxies \citep[see][for more details]{Wang2015,Blank2019}.

The NIHAO simulations have been proven to be very successful in reproducing several observed
scaling relations like the Stellar Halo-Mass relation \citep{Wang2015}, the disc gas mass and disc size relation \citep{Maccio2016}, the Tully-Fisher relation \citep{Dutton2017}, and the diversity of galaxy rotation curves \citep{Santos-Santos2018}. The above results show that NIHAO galaxies have a realistic 
distribution of the luminous and total mass and hence constitute an ideal set of simulations 
to study the distribution of  dark matter in galaxies.
A subset of NIHAO simulations \citep{Wang2015} has been already analyzed in the \citetalias{Tollet2016}
paper, with respect to that original work we have here added 29 galaxies created with the same
code and choice of parameters for the subgrid physics.

In order to perform simulations of massive galaxies able to show realistic stellar masses and star formation rates, we have recently added to the {\sc gasoline2} code a simple model for Black Hole formation, accretion and feedback. The model is based on Bondi–Hoyle–Lyttleton accretion, limited by the Eddington rate,
while the feedback is the form of pure thermal energy (no momentum) distributed isotropically to the gas surrounding the Black Hole \citep[see][for a detailed discussion and a thorough parameter study]{Blank2019}.
We have a total of 46 new simulations, in the stellar mass range $10^{10}-10^{12}$ \Msun, 
all galaxies are resolved with more than half million elements and are all central galaxies, like
in the original NIHAO sample.

\subsection{Profile Fit}

With the aim of maintaining a continuity with previous works, we strictly follow the methodology described in \citetalias{Tollet2016}.
We determined the halo center by using the shrinking sphere method \citep{Power2003} and then we divided the halo in fifty spherical shells with constant width on logarithmic scale.
For each shell we evaluated the average Dark Matter density obtaining the halo density profile.

Lastly, we considered the shells with radius between 1\% and 2\% of the virial radius,
and we fitted the density profile with a linear fit in the $\log r$-$\log \rho$ scale to obtain the profile slope $\alpha$ ($\rho\propto r^{-\alpha}$). The virial Radius has been set as the radius where the halo overdensity is equal to 200 times the critical density of the universe. 
We have (re)computed the value of the inner slope $\alpha$ for all galaxies
regardless if they were already been analyzed in  \citetalias{Tollet2016}.

%%%%%%%%%%%%%%%%%%%%%%%%%%%%%%%%%%%%%%%%%%%%%%%%%%%
\section{Results}\label{sec:results}

%%%%%%%%%%%%%%%%%%%%%%%%%%%%%%%%%%%%%%%%%%%%%%%%%%

\begin{table*}\label{table}
\begin{tabular}{lccccccccc}
\hline
$x$  [M$_{\odot}$] & $n$ & $n_1$ & $x_0$ [M$_{\odot}$] &  $x_1$ [M$_{\odot}$] &  $x_2$ [M$_{\odot}$] & $\beta$ & $\gamma$ & $\delta$ & 1-$\sigma$\\
 \hline
$M_{\rm halo}$ & -1.35 & 12.64 & $9.24\cdot 10^{11}$  & $9.10\cdot 10^{8}$  & $2.72\cdot 10^{12}$ & 1.27 & 1.14 & 1.68 & 0.32\\
$M_{\rm star}$ & -1.89 & 0.532 & $1.33\cdot 10^{11}$  & $1.23\cdot 10^{5}$  & $1.33\cdot 10^{11}$ & 0.419 & 0.826 & 1.52 & 0.28\\
\hline
\hline
$x$ & $n$ & $n_1$ & $n_2$ & $x_0$ &  $x_1$  & $\beta$ & $\gamma$ &  $\epsilon$ &  1-$\sigma$\\
 \hline
 $M_\mathrm{stars}/M_\mathrm{halo}$ & -0.0385 & 39.11 & -2.58 & $7.51\cdot 10^{-3}$  & $5.12\cdot 10^{-5}$  & 0.728 & 1.84 &  0.708 & 0.37 and
0.28\\
\hline
\end{tabular}
\caption{Parameters values for the fitting functions described in eq. \ref{eq1} and \ref{eq2}.}
\end{table*}

%%%%%%%%%%%%%%%%%%%%%%%%%%%%%%%%%%%%%%%%%%%%%%%%%%

\begin{figure*}
\includegraphics[width=0.9\textwidth]{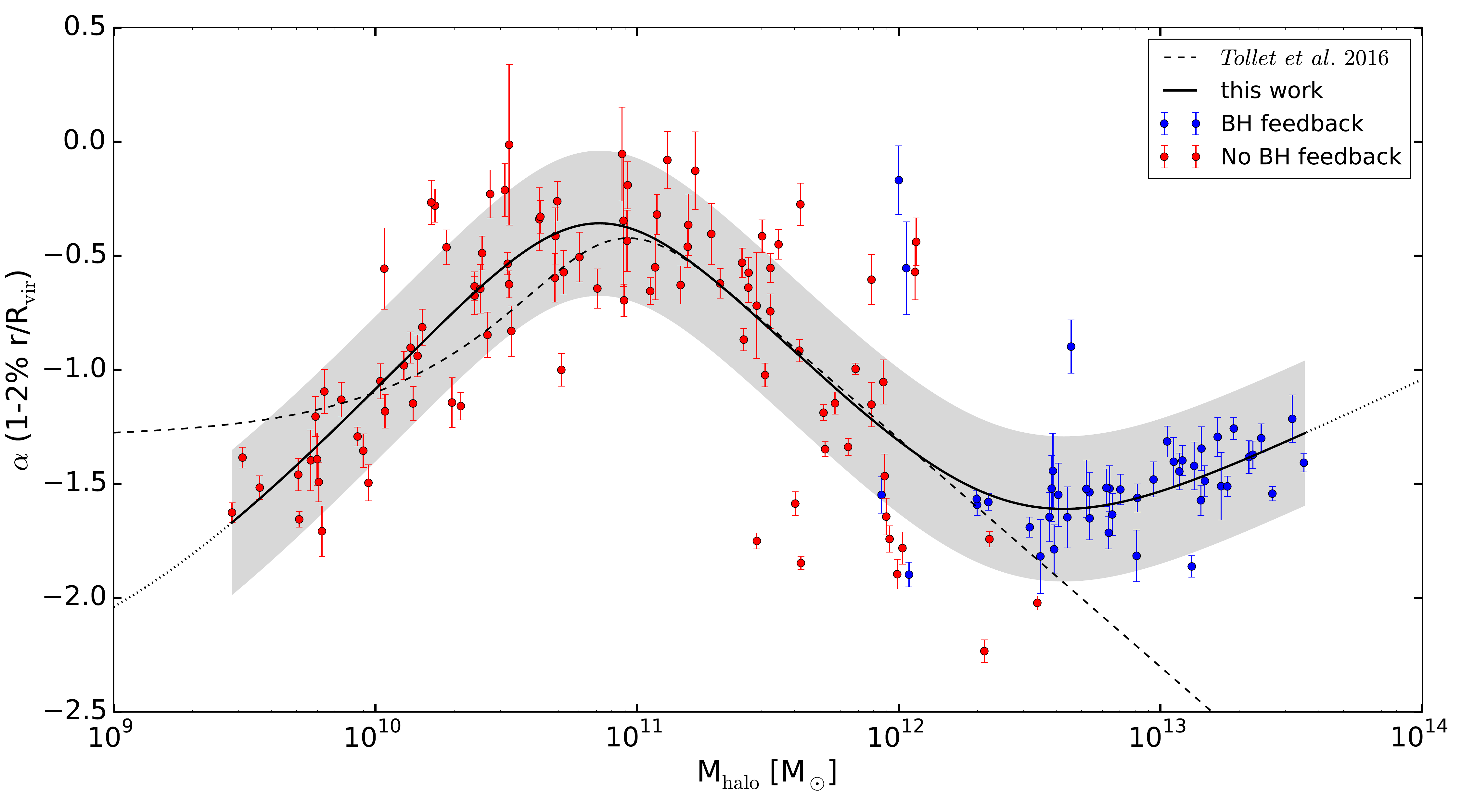}
\vspace{-.4cm}
\caption{Halo inner density profile slope $\alpha$ as a function of the halo mass $M_{\rm halo}$ at redshift $z=0$. The red and blue symbols represent galaxies without and with BH feedback, respectively.
The black solid line corresponds to the best fit from eq. \ref{eq1} of the combined catalog; the dashed line is the relation proposed in  \citetalias{Tollet2016}, based on a subset of the NIHAO galaxies without BH.
The gray shaded area shows the one sigma scatter (0.32) for the whole sample: the scatter in the mass bins $10^{10}$, $10^{11}$, $10^{12}$, $10^{13}$ \Msun with a $\Delta M_{\rm halo}$ of one dex is: 0.26, 0.28, 0.50, 0.18 respectively.}
\label{fig:Mhalo}
\end{figure*}

\begin{figure*}
\includegraphics[width=0.9\textwidth]{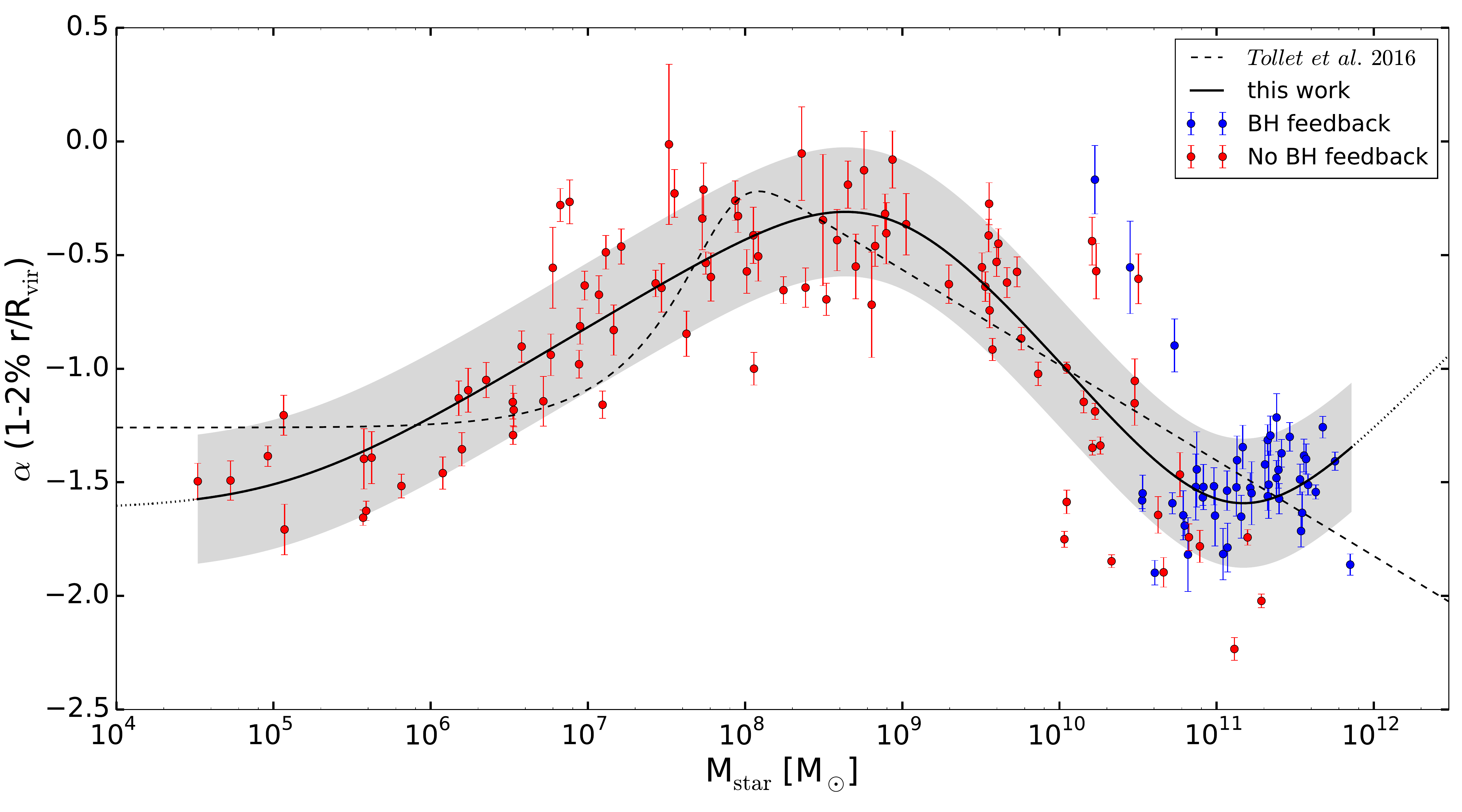}
\vspace{-0.4cm}
\caption{Halo inner density profile slope $\alpha$ as a function of the stellar mass $M_{\rm star}$.
Colors, symbols and fitting functions are the same as in Fig. \ref{fig:Mhalo}.
The global one sigma scatter represented by the gray area is 0.28, while scatter in the mass bins $10^{5}$, $10^7$, $10^{9}$, $10^{11}$ \Msun with a $\Delta M_{\rm star}$ of two dex is: 0.19, 0.26, 0.22, 0.33 respectively.}
\label{fig:Mstar}
\end{figure*}

Our main results are summarized by Figures  \ref{fig:Mhalo} and \ref{fig:Mstar}, where we show the dependence of the inner density slope $\alpha$ as a function of the halo and stellar mass respectively.
In all plots red symbols are for simulations without AGN feedback \citetalias[an extended sample w.r.t. ][]{Tollet2016}), while blue ones are for simulations including AGN feedback. 

In the halo mass range $2\times 10^9-10^{12} \Msun$ we confirm the previous findings from  \citetalias{Tollet2016}, namely a non-monotonic behaviour of the halo response to galaxy formation, with a peak of ``core-creation" for $M_{\rm halo}\approx 10^{11} \Msun$. In the new explored halo mass 
range ($M_{\rm halo}>2 \times 10^{12})$ the new simulations including AGN feedback deviate from a mere extrapolation of the \citetalias{Tollet2016} fitting formula and show a progressive relaxation of the halo with a density profile shallower than that predicted by pure Nbody simulations \citep{dutton_maccio14}, even though still cuspy.

The same considerations apply to the behaviour of $\alpha$ vs the stellar mass shown in Figure \ref{fig:Mstar}, with the new simulations with AGN feedback showing an up-turn in the relation for 
$M_{\rm star}>10^{11} \Msun$.

We tried to capture the overall behavior of the halo inner slope as a function of $M_\mathrm{halo}$ and $M_\mathrm{star}$  using an updated version of the fitting function proposed in \citetalias{Tollet2016}:
\begin{equation}\label{eq1}
\alpha(x)=n-\log_{10}\left[n_1\left(1+\frac{x}{x_1}\right)^{-\beta}+\left(\frac{x}{x_0}\right)^\gamma\right]
+\log_{10}\left[1+\left(\frac{x}{x_2}\right)^\delta\right]\, .
\end{equation}
where the only difference is the extra additive term.
We use this equation for fitting the combined galaxy catalogue; results are listed in table \ref{table}.

%%%%%%%%%%%%%%%%%%%%%%
%%%%%%%%%%%%%%%%%%%%%%

%%%%%%%%%%%%%%%%%%%%%%
%%%%%%%%%%%%%%%%%%%%%%

\begin{figure}
\includegraphics[width=0.5\textwidth]{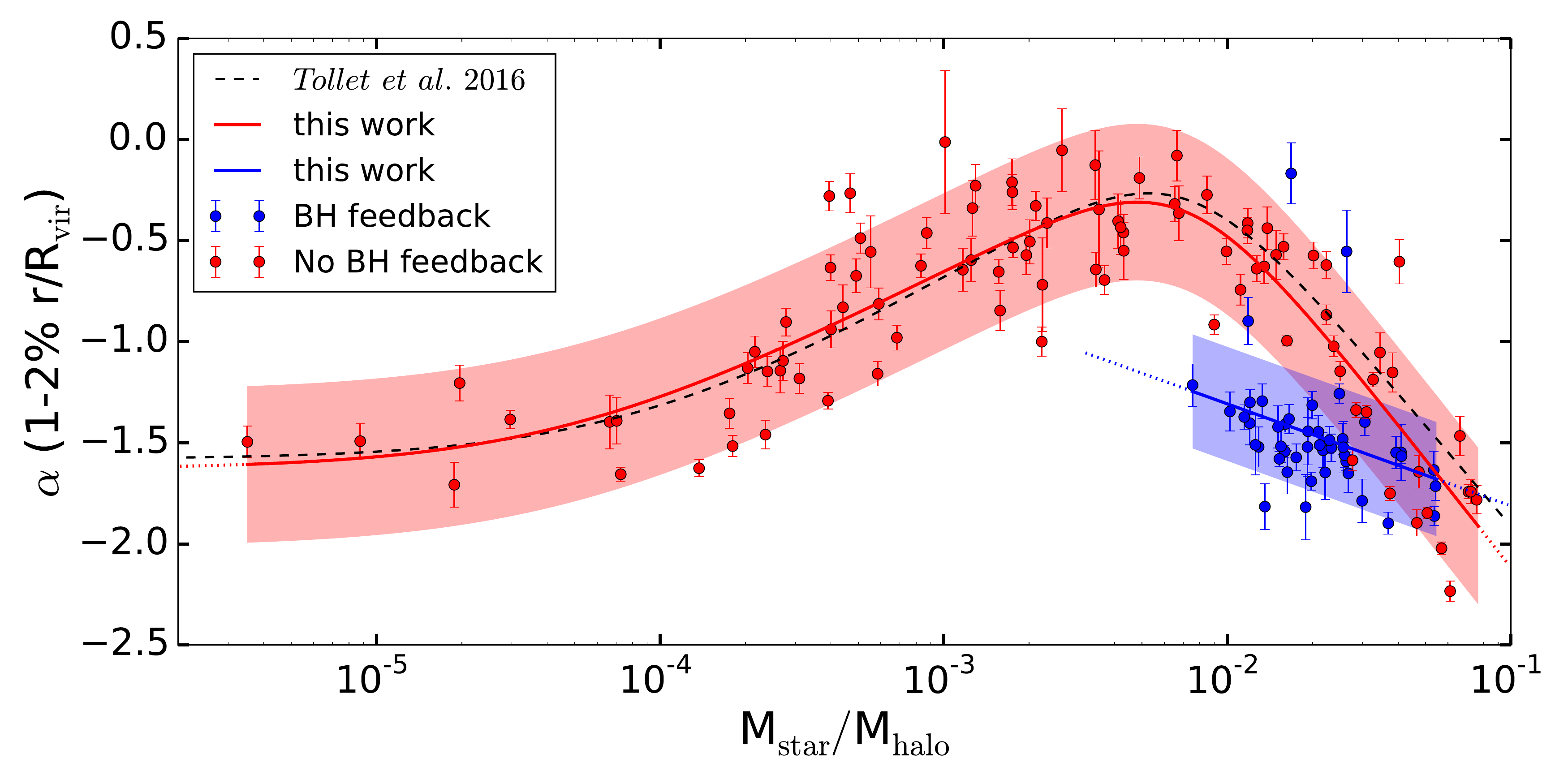}
\vspace{-0.4cm}
\caption{Halo inner density slope $\alpha$ as a function of the star formation efficiency  $M_\mathrm{star}/M_\mathrm{halo}$ at redshift: $z=0$. Blue and red symbols represent galaxies with 
and without BH feedback respectively. The red solid line corresponds to the best fit (eq. \ref{eq2}, first branch) of galaxies without BH and the red area is the one sigma scatter (0.37); the blue solid line corresponds to the best fit (eq. \ref{eq2}, second branch) for galaxies with BH feedback, the blue area is the one sigma scatter (0.28). The dashed line is the original fit from \citetalias{Tollet2016}.}
\label{fig:Mstarhalo}
\end{figure}

In previous works \citep[e.g.][]{DiCintio2014a, Chan2015} it has been shown that the integrated
star formation rate, parametrized as the ratio between $M_{\rm star}/M_{\rm halo}$, is the actual driver that describes the changes in the dark matter distribution. 
The results for our simulations are shown in Figure  \ref{fig:Mstarhalo}; since the relation between the star formation efficiency and the halo mass is not monotonic, massive galaxies somehow ``walk backwards" in the plot and are over-imposed to lower mass galaxies, which limits the use
of this parameter to describe the whole population.
To mathematically describe the $\alpha$ 
dependence on $M_\mathrm{star}/M_\mathrm{halo}$ we use a combination of \citetalias{Tollet2016} original fitting function and a linear fit in the semilog scale:
\begin{equation}\label{eq2}
\alpha\left(x\right)=
\begin{cases}
n-\log_{10}\left[n_1\left(1+\frac{x}{x_1}\right)^{-\beta}+\left(\frac{x}{x_0}\right)^\gamma\right], & \mbox{if } M_\mathrm{halo} \lesssim 4\cdot 10^{12} \mbox{M}_\odot \\
n_2-\epsilon\log_{10}x, & \mbox{if } M_\mathrm{halo} \gtrsim 4\cdot 10^{12} \mbox{M}_\odot
 \end{cases}\, .
\end{equation}
Since the two branches of this relation are obtained by fitting two different catalogs, the turning point must be used with caution. This point reasonably correspond to the minimum of $\alpha(M_\mathrm{halo})$, that lays around $M_\mathrm{halo} \sim 4\cdot 10^{12} \mbox{M}_\odot$, or equivalently 
$M_\mathrm{star} \sim 2\cdot 10^{11} \mbox{M}_\odot$.

The scatter around all the proposed relations is constant with a value of about $\sigma=0.32$ for the $\alpha-M_{\rm halo}$ relation and of 0.28 for the $\alpha-M_{\rm star}$ one (this scatter is computed using all galaxies in the sample, a mass dependent value of the scatter is reported in the caption of figures \ref{fig:Mhalo} and \ref{fig:Mstar}).
A constant scatter is a good approximation for most of the mass range inspected, with the exception of a region around $10^{12}$ ($2\times10^{10}$) \Msun in halo  (stellar) mass. This larger scatter seems to be independent of the presence or absence of AGN feedback, and instead more inherent to this specific mass scale. 

\begin{figure}
\includegraphics[width=0.46\textwidth]{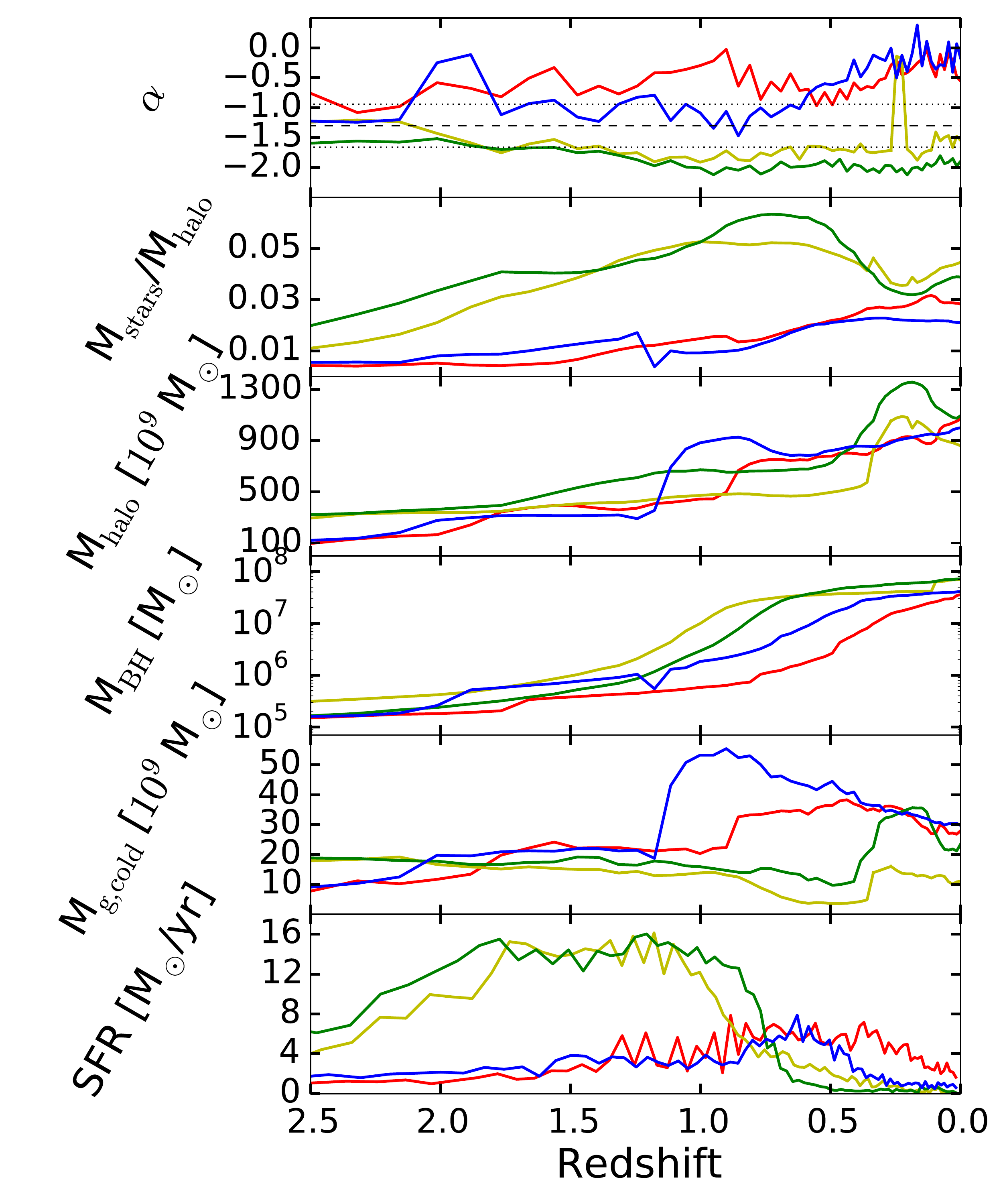}
\vspace{-0.4cm}
\caption{
Time evolution for different parameters for four halos with halo mass $M_\mathrm{halo}\sim 10^{12}$ M$_\odot$ at $z=0$. 
From top to bottom: DM inner slope $\alpha$, $M_\mathrm{star}/M_\mathrm{halo}$ ratio, halo mass $M_\mathrm{halo}$, black hole mass $M_\mathrm{BH}$, cold gas mass $M^\mathrm{g,cold}$ and star formation rate.
The dashed line in the top panel shows the average value for $\alpha$ for a halo mass $\sim 10^{12}$  M$_\odot$ at z=0; the dotted lines indicate the 1-$\sigma$ scatter.}
\label{fig:weirds}
\end{figure}

In order to better understand the origin of this large scatter we have looked in details to four galaxies, with similar halo masses at $z=0$ ($M_\mathrm{halo} \sim 10^{12} \mbox{M}_\odot$) but very different values for $\alpha$: two with shallow
density profiles ($\alpha>-0.5$), and two quite cuspy $(\alpha<-1.5)$. 
In Figure  \ref{fig:weirds} we show the time evolution of five different parameters that might influence the DM reaction to galaxy formation: namely, the black hole mass, the amount of cold gas, the stellar to mass ratio, the halo mass, and the star formation rate. 

Our analysis suggests that the main driver for the different $\alpha$ of these galaxies is the different assembly time of the stellar mass. The two galaxies  with cuspy profiles at $z=0$ have a vigorous SFR in between $z=1$ and $z=2$, as a consequence their stellar to halo mass ratio is very high
in the past, which causes the halo to contract, due to its large stellar body (see the discussion in \citetalias{Tollet2016}). Such a contracted halo is set very early ($z\approx2$) and, once
established, it is very hard for outflows generated by the stellar or AGN feedback to reverse it. 
On the contrary the two galaxies with shallow profiles at $z=0$ have a less intense SFR, as a consequence the stellar body forms more gradually, and the potential fluctuations due to gas outflows
are able to expand the dark matter halo distribution \citep{Pontzen2012}.

\section{Discussion and Conclusions}
\label{sec:conclusions}

In this Letter we have presented a systematicc analysis of the reaction of the dark matter distribution to galaxy formation on more than eight orders of magnitute in stellar mass.
This work is based on the extended NIHAO suite \citep{Wang2015}, that now includes AGN feedback for
 massive haloes \citep{Blank2019} and counts more than 140 high resolutions galaxies, with halo masses from $2 \cdot 10^9 \Msun$ till $5\cdot 10^{13} \Msun$.
 
 We confirm the results previously obtained on a restricted mass range by \citetalias{Tollet2016} \citep[see also][]{DiCintio2014a,Chan2015}, namely that the halo response is not monotonic,
 low mass haloes ($M_{\rm halo}<2\cdot 10^{10}$\Msun) retain the original cuspy CDM profile, then there is a progressive relaxation of the halo that culminates with the creation of a dark matter core
 for $M_{\rm halo}\approx 10^{11}\mbox{M}_\odot$, after this pivot point the halo starts to progressively contract back reaching a slope of -1.5 at $M_{\rm halo}\approx 10^{12} \mbox{M}_\odot$. 
 
 For the first time we have explored the halo reaction on mass scales beyond 
 $10^{12}\mbox{M}_\odot$ with a large number of galaxies. We found that  outflows  generated  by  AGNs are  able  to  partially counteract the dark matter contraction on these scales, and there is a relaxation of the halo, 
 with an increasing slope of the density profile, that sits around -1.2 at the edge of our mass range. 
 The behaviour of the inner density slope of the dark matter distribution, $\alpha$, as a function of the halo or stellar mass is captured by a simple analytical behaviour. 
 The scatter around the mean value is constant and of the order of 0.32, with the exception of haloes
 around $10^{12} \mbox{M}_\odot$. At this mass scale, where we see the peak of star formation efficiency, the competing behaviour of contraction due to the creation of a large stellar body, and the expansion due to vigorous outflows (due to both stellar and AGN feedback), more than doubles the scatter and hence any mean expectation value, especially if compared with observations, should be used with extreme care. 
 
 Finally we provide easy to use fitting formulas that characterize the halo response as a function of 
 stellar and halo mass; such formulas can be of use to set up dark matter haloes taking into account the effect of galaxy formation and to make comparisons with observed values.

\section*{Acknowledgements}

The authors thank the referee of this papaer, Alan R. Duffy for his comments that improved the readability of our work.
The authors gratefully acknowledge the Gauss Centre for Super-computing e.V. (www.gauss-centre.eu) for funding this project by providing computing time on the GCS Supercomputer SuperMUC at Leibniz Supercomputing Centre (http://www.lrz.de). A part of this research was carried out on the High Performance Computing resources at New York University Abu Dhabi. We used the software package PYNBODY \citep{pontzen_etal13} for  our  analyses. 

%%%%%%%%%%%%%%%%%%%% REFERENCES %%%%%%%%%%%%%%%%%%

% The best way to enter references is to use BibTeX:

\bibliographystyle{mnras}
\bibliography{ref} % if your bibtex file is called example.bib

% Alternatively you could enter them by hand, like this:
% This method is tedious and prone to error if you have lots of references

%%%%%%%%%%%%%%%%%%%%%%%%%%%%%%%%%%%%%%%%%%%%%%%%%%

%%%%%%%%%%%%%%%%% APPENDICES %%%%%%%%%%%%%%%%%%%%%

%\appendix

%%%%%%%%%%%%%%%%%%%%%%
%%%%%%%%%%%%%%%%%%%%%%

%%%%%%%%%%%%%%%%%%%%%%
%%%%%%%%%%%%%%%%%%%%%%

% Don't change these lines
\bsp	% typesetting comment
\label{lastpage}
\end{document}